\documentclass[aps,prb,twocolumn,showpacs,groupedaddress]{revtex4}
\usepackage{dcolumn}% Align table columns on decimal point
\usepackage{bm}% bold math
\usepackage{graphicx}
\begin{document}

\newcommand{\lin}{PbCuSO$_4$(OH)$_2$}

%\preprint{APS/123-QED}

\title{The dynamics of linarite: Observations of magnetic excitations}
\author{K. C. Rule$^1$, B. Willenberg$^{2,3}$, M. Sch\"{a}pers$^{4}$, A. U. B. Wolter$^4$, B. B\"{u}chner$^4$, S.-L. Drechsler$^4$, G. Ehlers$^5$, D. A. Tennant$^5$, R. A. Mole$^1$, J. S. Gardner$^{6,7}$, S. S\"{u}llow$^2$, and S. Nishimoto$^{4,8}$}

\affiliation{$^1$Australian Nuclear Science and Technology Organisation, Kirrawee DC, NSW 2234, Australia \\
$^2$Institute for Condensed Matter Physics, TU Braunschweig, 38106 Braunschweig, Germany \\
$^3$Helmholtz-Zentrum Berlin f\"ur Materialien und Energie, 14109 Berlin, Germany \\
$^4$Leibniz Institute for Solid State and Materials Research, IFW Dresden, 01171 Dresden, Germany \\
$^5$Quantum Condensed Matter Division, Oak Ridge National Laboratory, Oak Ridge, Tennessee 37831, USA\\
$^6$Neutron Group, NSRRC, 101 Hsin Ann Road, Hsinchu, 30076 Taiwan\\
$^7$Center for Condensed Matter Sciences, National Taiwan University, Taipei 10617, Taiwan\\
$^8$Department of Physics, Technical University Dresden 01069 Dresden, Germany} 

\begin{abstract}
Here we present inelastic neutron scattering measurements from the frustrated, quantum spin-1/2 chain material linarite, \lin. Time of flight data, taken at 0.5\,K and zero applied magnetic field reveals low-energy dispersive spin wave excitations below 1.5\,meV both parallel and perpendicular to the Cu-chain direction. From this we confirm that the interchain couplings within linarite are around 10\,\% of the nearest neighbour intrachain interactions. We analyse the data within both linear spin-wave theory and density matrix renormalisation group theories and establish the main magnetic exchange interactions and the simplest realistic Hamiltonian for this material.
\end{abstract}

\maketitle

Fundamental research into low-dimensional and frustrated magnets has gained momentum recently due to advances in computational power and experimental measurement techniques. The combination of low dimensionsality, frustration and quantum physics effectively suppresses conventional long range order down to very low temperatures, which can lead to unconventional magnetic states such as quantum spin liquids \citep{Balents2010,Helton2007}, spin-Peierls states \citep{Arai1996} and Tomonaga-Luttinger liquid phases \citep{WillenCuN}. Additionally, when a magnetic field is applied to such systems, a range of exotic states such as spin-multipolar phases \citep{Chubukov1991,Vekua2007,ZhitomirskyLiCuVO4,Nawa2013LiCuVO4,Buttgen2014} may be induced.

Especially, the so-called spin-nematic state has recently received strong interest \citep{Buttgen2014,Mourigal2012,Starykh2014,Nawa2014NaCu,Onishi2015,Du2016}. The spin-nematic phase can be likened to the arrangement of molecules in nematic liquid crystals. The state involves the ordering of spin-multipole moments without conventional spin-dipole order such that the magnetic spins align spontaneously along a chosen axis while still fluctuating dynamically. More formally we can say that a spin-nematic state breaks the spin rotational symmetry while preserving translational and time reversal symmetries, in contrast to conventionally ordered magnets \citep{Chubukov1991,Buttgen2014,Andreev1984}.   

Among others a magnetic model predicted to host the spin-nematic phase is a $J_1$-$J_2$ spin chain with competing ferromagnetic (FM) nearest-neighbor (NN) ($J_1 < 0$) and antiferromagnetic (AFM) next-nearest-neighbour (NNN) interactions ($J_2 > 0$). Further, the diagonal interchain coupling plays an important role, although the details are not yet fully understood: some coupling is required to establish long-range order, but above a small critical interchain coupling it can also destroy any isotropic multipolar phase \citep{Buttgen2014,Nishimoto2015}. However, with specific exchange anisotropies a stabilization of the spin-nematic phase can be achieved \citep{WillPRL2016,Grafe2016}. Together, the Hamiltonian, including diagonal interchain coupling, $J_{ic}$, between chains $l, l'$, is written as
\begin{equation}
\hat{H} = \sum_{l,m=1} ^{m=2}J_m{\bf S}_l \cdot {\bf S}_{l+m} + J_{ic} \sum_{l,l'} {\bf S}_{l} \cdot {\bf S}_{l'} - h \sum_{l} S_l^z. 
\label{Hamiltonian}
\end{equation}
Here, $S_{l}$ is a spin-1/2 operator on chain site $l$, and $h$ an external magnetic field. The level of frustration in case of antiferromagnetic NNN exchange $J_2$ is measured by the parameter $\alpha = J_2/\left| J_1 \right|$, which serves as indicator of the classical and quantum magnetic ground state. In the isotropic exchange case for $0 < \alpha < 0.25$ a FM ground state occurs, while for larger $\alpha$-values  the ground state is given by a non-collinear spin-spiral, or a singlet ground state occurs in the 1D and chiral correlations in the quantum case. The magnetic phase diagram of these materials is expected to be exotic where the spin correlations may change significantly in applied fields due to the ferromagnetic $J_1$, stabilizing bound magnon pairs. These bound magnon pairs form a spin density wave (SDW) in moderate fields, whereas, in fields just below the saturation magnetization they can exhibit Bose condensation into multipolar states \citep{Sudan2009,Furukawa2010,Sato2013}. One of them expected just below saturation is the quadrupolar state of bound magnon pairs which is termed the spin-nematic state \citep{Sudan2009}.

\begin{figure}
\begin{center}
\includegraphics[width=1.0\linewidth]{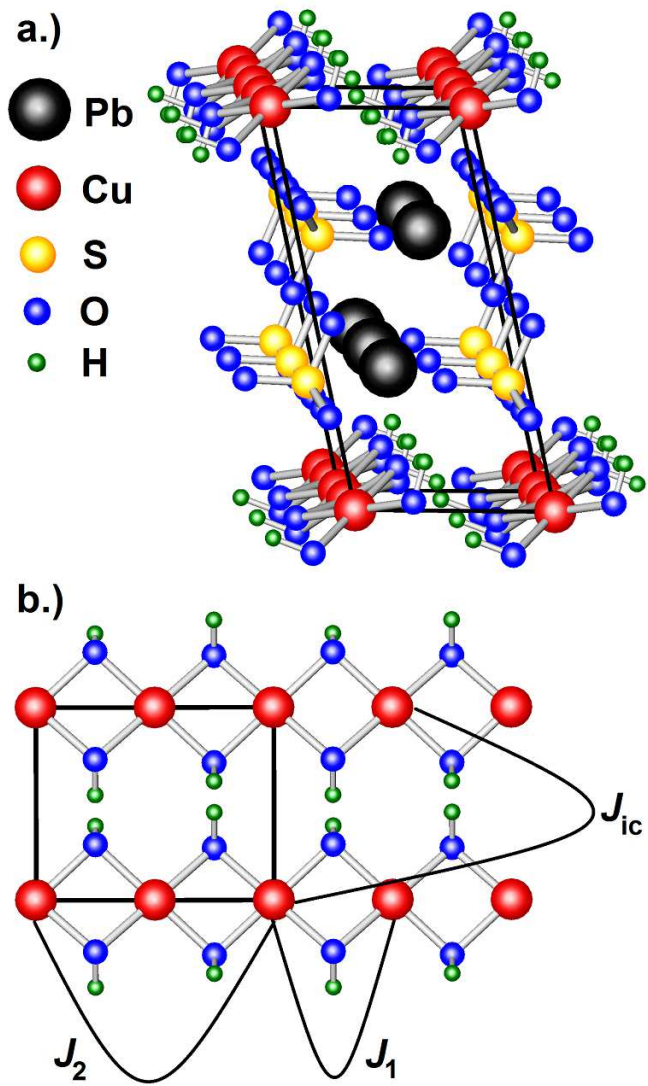}
\end{center}
\caption{(Color online) The crystal structure of linarite, with a.) a view onto the $a$-$c$ and b.) the $b$-$c$ plane, with the Cu$^{2+}$ chain structure directed along the $b$ axis. The exchange couplings $J_1$, $J_2$ and $J_{ic}$ are indicated; for details see text.}
\label{fig:structure}
\end{figure}

The $J_1$-$J_2$ model has been used to describe various quasi-1D edge-sharing cuprates \citep{Nawa2014NaCu,Sato2013,Kamieniarz2002,Hase2004,Masuda2004,Masuda2005,Enderle2005,Drechsler2006,Svistov2011,Yasui2011}. Unfortunately, for many of the candidates for exhibiting multipolar phases, the involved AFM magnetic couplings and thus the saturation field are quite large, making it difficult to probe the multipolar phases. In this respect, linarite, \lin, is a promising spin-multipolar candidate as the Cu$^{2+}$ ions form a quasi-1D $S = 1/2$ spin-chain along the crystallographic $b$-axis \citep{WillPRL2016,Wolter2012,WillPRL2012,Schapers2013,Schapers2014}. The material crystallizes in a monoclinic space group $P2_1/m$ with lattice parameters \cite{WillPRL2012,Schapers2013,Effenberger1987} (at 1.8\,K) $a = 9.682$\,\AA, $b = 5.646$\,\AA, $c = 4.683$\,\AA, and $\beta$ = 102.65$^{\circ}$ (Fig.~\ref{fig:structure}). Since linarite has a small saturation field $\sim$\,10\,T, and is available in natural single crystal form, it is considered as an ideal material for studies of the multipolar phases.

We have recently presented an extensive thermodynamic investigation of linarite, establishing a rich and detailed phase diagram along each of the principal directions \citep{WillPRL2016,WillPRL2012,Schapers2013, Schapers2014}. Below $T_N = 2.8$\,K linarite shows magnetic long range order at zero field. Neutron scattering experiments \citep{WillPRL2012} have confirmed the spin structure of the ground state to be an elliptical helix with the propagation vector $\vec{{\bf k}} = (0,0.186,0.5)$. From bulk magnetic measurements, we have estimated the magnetic exchange interactions, with FM-NN $J_1 \approx -100$\,K, AFM-NNN $J_2 \approx 36$\,K and $J_{ic} \sim 5$\,K \citep{Wolter2012}. This gives two coupled $J_1$-$J_2$ chains with $\alpha \approx 0.36$ and $J_{ic} = 0.05 |{J_1}|$. For linarite this interchain coupling was revealed to be a critical component into the multipolar ordering at high magnetic fields \citep{WillPRL2016}. While the incommensurate propagation vector component (in units $[H,K,L]$) along the chain has been found to be $K = 0.186$ in zero field, for a single $J_1$-$J_2$ chain with $\alpha \approx 0.36$ a value $K = 0.367$ would be predicted. This discrepancy can be resolved by taking into account the residual interchain interaction.

Inelastic neutron scattering (INS) is the most direct probe for investigating the magnetic excitations in low dimensional quantum magnets \citep{azuriteFLEX, azuriteOSIRIS}. In order to refine the magnetic exchange interactions in \lin, here we present INS data revealing the zero field excitation spectra of linarite. We model our data by means of linear spin-wave theory (LSWT) as well as dynamical density-matrix renormalization group (DDMRG) analysis and dicuss merits and limits of these approaches.

For frustrated CuO$_2$-chain compounds with FM $J_1$-values, to the best of our knowledge only four materials have been studied by INS: LiCu$_2$O$_2$ \cite{Masuda2005}, LiVCuO$_4$ \cite{Enderle2005}, Li$_2$CuO$_4$ \cite{Lorenz2009} and Ca$_2$Y$_2$Cu$_5$O$_{10}$ \cite{Kuzian2011}. The first two exhibit non-collinear incommensurate spiral order like linarite, while the last two show commensurate collinear AFM order with FM aligned chains. In the latter cases LSWT provides a description in accord with DMRG-calculations. In contrast, for the first two materials significant discrepancies remain concerning the magnitude of the interchain coupling in case of LiCu$_2$O$_2$ \cite{Gippius2004,Drechsler2005,Mazurenko2007} and the $\alpha$-value in case of LiVCuO$_4$ \cite{Drechsler2011,Nishimoto2012,Sirker2010,Ren2012}. We ascribe these deviations mainly to the crucial role of quantum fluctuations beyond the LSWT. It remains to be seen to what extent a recently proposed approach beyond LSWT \cite{Du2016} will improve the situation for such spiral materials.

Due to the small sample size of the naturally grown linarite crystals, a multi-crystal array was prepared for INS measurements by co-aligning 5 needle shaped crystals on copper wafers. The oxygen free copper wafers were chosen to ensure good thermal contact with the sample at temperatures below 1\,K. The samples were aligned in the $[0,K,L]$ plane and the resulting quasi-single crystal had a mosaic spread of less than 3$^{\circ}$. Experiments were carried out at Oak Ridge National Laboratory using the Cold Neutron Chopper Spectrometer in conjunction with a $^3$He-insert \citep{CNCS}. The sample was aligned within the cryostat and complete data sets were measured at 0.5 and 40\,K ({\it i.e.}, well below and above $T_N$, with the highest temperature chosen to be just larger than $J_2$). Since two different energy scales were expected from the incommensurate spin-wave calculations, the instrument was tuned to provide incident energies of $E_i = 3.315$\,meV (4.967\,\AA ) and 12.03\,meV (2.611\,\AA ) which afforded energy resolutions at the elastic line of 0.05 and 0.5\,meV, respectively. This way, in particular the low energy excitations could be observed with high resolution. For each incident energy and temperature the sample was rotated through 180$^{\circ}$ in 2$^{\circ}$ steps. The data were then combined using MSLICE to produce a data set over the complete S(Q,$\omega$) range which could be sliced to extract anisotropic scattering data \citep{mslice}.

\begin{figure}
\begin{center}
\includegraphics[width=1.0\linewidth]{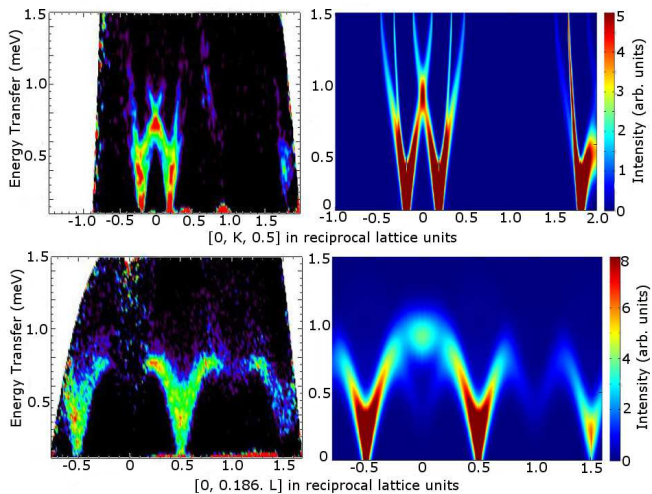}
\end{center}
\caption{(Color online) Measured (left) and calculated (LSWT, right) spin wave excitations of linarite at 0.5\,K for $E_i$ = 3.315\,meV. The upper row shows the interactions along the Cu-chain $[0,K,0.5]$, the lower row those perpendicular to the chain $[0,0.186,L]$ ({\it i.e.}, $\parallel J_{ic}$). Data have been corrected for background scattering by subtracting the 40\,K spectra.}
\label{fig:LowE}
\end{figure}

The low temperature INS results, taken with low incident energy neutrons, are displayed in Fig.~\ref{fig:LowE}, along with the corresponding LSWT calculations. The measured data are shown in the left columns and have been corrected for background scattering. Since the data at 40\,K appeared to have no low-energy, magnetic spin-wave excitations, this data set was used to remove the background scattering contributions from the sample environment and mount from the weak magnetic scattering of the sample. These results have also been corrected for detector efficiency by applying a vanadium measurement during data processing. Fig.~\ref{fig:LowE} focuses on the low energy scattering both parallel (top) and perpendicular (bottom) to the Cu-chains in linarite. The scattering profile for the excitations along the Cu-chain (Fig.~\ref{fig:LowE} upper left) shows multiple modes which collapse to the elastic line at the incommensurate wave-vector component $K = \pm 0.186$ as expected. There is a crossover of these modes at around 0.75\,meV at $K = 0$.

Even in this low energy regime, it is clear that these modes extend to much higher energies with a steep slope, and which we tried to study in the high incident energy experiments. Unfortunately, due to a combination of an extremely small sample, small Cu$^{2+}$ moments, scattering from the sample mount and reduced scattering intensities, we were not able to discern any distinct features from the background scattering beyond an energy transfer of around 1.5\,meV.

For a truly 1D system, with $J_{ic} = 0$, one would expect to see no dispersive excitations perpendicular to the chain. Thus the data in Fig.~\ref{fig:LowE} (lower left) indicate the significant role that $J_{ic}$ plays in this system. Despite the frustrated NN and NNN coupling, it is the interchain coupling which allows for long range magnetic order below 2.8\,K. Actually, the data along $[0,0.186,L]$ show scattering reminiscent of a 1D Heisenberg AFM chain with regular arches reaching to around 0.8meV.  

To parametrize our data, LSWT calculations were performed using the packet SpinW \citep{SpinW}. It attempts to solve the spin Hamiltonian using both classical and quasi-classical numerical methods. In this software the calculated magnetic ground state of the long-range ordered system is used with the Holstein-Primakoff approximation for the spin operator to model the observed spin-wave excitations \citep{TothLake}. Since it incorporates incommensurate magnetic structures, it is suitable for calculations of the spin-wave excitations in linarite. In the given context, exchange interactions of $J_1 = -114$\,K (FM), $J_2 = 37$\,K (AFM), and $J_{ic} = 4$\,K (AFM) lead to a calculated magnetic structure with a propagation vector $\vec{{\bf k}} = (0,0.185,0.5)$, in good agreement with the experimentally observed one \citep{WillPRL2012}. 

\begin{figure}
\begin{center}
\includegraphics[width=1.0\linewidth]{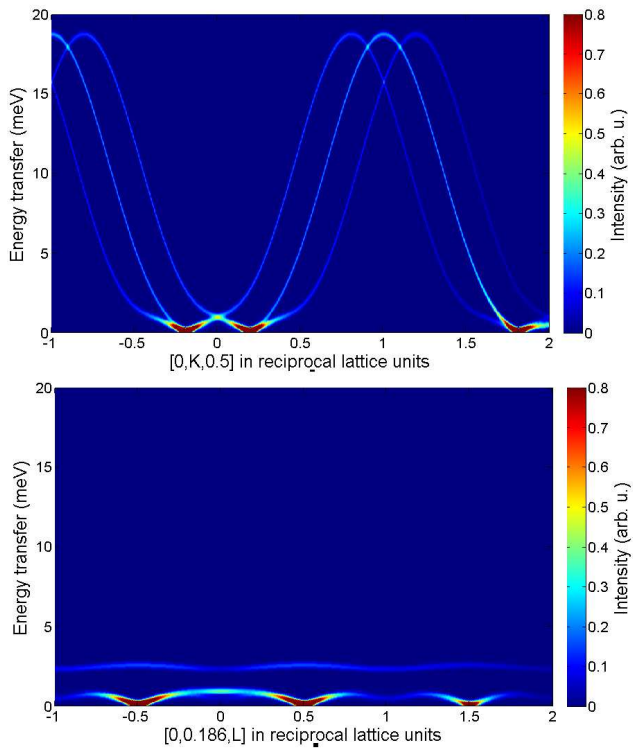}
\end{center}
\caption{(Color online) Calculated spin wave excitations (LSWT) of linarite along the Cu-chain direction $[0,K,0.5]$ (upper panel) and perpendicular to the chain along $[0,0.186,L]$ (lower panel).}
\label{fig:highE}
\end{figure}

In turn, these calculations also resulted in the calculated spin-wave spectra shown in the Figs.~\ref{fig:LowE} and \ref{fig:highE}. Clearly, these calculations reproduce the salient features of the low energy branch of the experimental spin excitation spectra. Thus, relatively good agreement is found when modeling the data with the exchange interactions $J_1$, $J_2$ and $J_{ic}$, as determined from susceptibility $\chi(T)$, magnetization $M(H)$ and inelastic neutron diffraction INS, indicating that the model used to describe the magnetic coupling in linarite is robust \citep{WillPRL2016, Wolter2012}, see Tab. \ref{tab:j}. 

\begin{table}[b]
\caption{Comparison of the main isotropic exchange integrals and the frustration ratio $\alpha$ for linarite as determined from the present INS studies and previous works.}
\label{tab:j}
\begin{tabular}{lcccc}
\hline
Technique & $J_1$ [K] & $J_2$ [K] & $J_{\rm ic}$ [K] & $\alpha$ \\ \hline
INS (LSWT) & $-114 \pm 2$ & $37 \pm 1$ & $4 \pm 0.5$ & 0.32 \\
INS (DDMRG) &  -78 & 28 & 7& 0.36 \\
$\chi(T)$, Ref. [\onlinecite{Wolter2012}] & -97.5 & 35.1 & -8.8 & 0.36 \\
$M(H)$, Ref. [\onlinecite{Wolter2012}] & -89.5 & 32.7 & -- & 0.37 \\
L(S)DA+U, Ref. [\onlinecite{Wolter2012}] & -133 & 42 & 7& 0.32 \\
$\chi(T)$, Ref. [\onlinecite{Yasui2011}] & -13 & 21 &  -- & 1.62 \\
$\chi(T)$, Refs. [\onlinecite{baran2006,Kamieniarz2002}] & $-30 \pm 5$ & 15 &  -- & $0.5 \pm 0.05$ \\
\hline
\end{tabular}
\end{table}

Note, that the discrepancy with in particular the analysis in Ref. [\onlinecite{Yasui2011}] remains striking. It suggests that fits only to high temperature susceptibility are insufficient to retrieve a precise set of $J$-parameters. Effectively, in Ref. [\onlinecite{Yasui2011}], the high-$T$ analysis employed well above the maximum of $\chi(T)$ provides no unique set of exchange parameters. In this context, an instructive example is LiCu$_2$O$_2$: in Ref. [\onlinecite{Masuda2004}], the authors arrived at an improper AFM $J_1$-value based on a high-$T$ susceptibility analysis, commented on in Ref. [\onlinecite{Drechsler2005}], and corrected later in Ref. [\onlinecite{Masuda2005}]. The maximum position of $\chi(T)$ affected by FM excitations above the spiral ground state is very sensitive to the distance of $\alpha$ to the critical point at 0.25 (see Ref. [\onlinecite{Drechsler2007}]). Hence, susceptibility fits need to include the low-$T$ region near the maximum of $\chi(T)$, as we did in Ref. [\onlinecite{Wolter2012}]. 

The excitations along the chain, {\it i.e.}, along the $[0,K,0.5]$ direction, are defined by the two energy scales of the $J_1$ and $J_2$ couplings. From the model, the $J_2$ AFM interaction appears to define the lower energy limits along the chain direction, at around 1\,meV, while the upper limits in this direction are defined by the $J_1$ interaction with spin-wave excitations expected to extend to around 20\,meV (Fig.~\ref{fig:highE}). The calculated spin-wave spectra also indicate that the scattering intensity of the high energy modes will be much weaker than the low energy part, consistent with our experiments. Perpendicular to the chain, {\it i.e.}, along $[0,0.186,L]$, there is no corresponding high energy scattering, reflecting the low-dimensional character of the spin excitations.

While LSWT describes well the excited states in an ordered magnet with large magnetic moments, it is known to break down for low dimensional magnetic structures and for quantum spins. As well, one limitation with LSWT is that there are no multi-magnon processes considered, which directly affect the fitting of the intensity at higher energies. Thus, while the data analysis using SpinW is a relatively simple means to estimate the expected magnetic excitations in linarite, it may not describe all of the interactions as a full quantum model. Therefore, in order to firmly establish the exchange interaction parameters, we fitted the low-energy excitations of the INS experiment using the dynamical density-matrix renormalization group (DDMRG) method~\cite{eric02}. We calculated the dynamical spin structure factor, which is defined as
\begin{equation}
S(Q,\omega) = \sum_\nu |\langle \psi_\nu |S^z_k| \psi_0 \rangle|^2 \delta(\omega-E_\nu+E_0),
\label{spec}
\end{equation}
where $S_i^z$ is the $z$-component of the spin-1/2 operator ${\bf S}_i$, $| \psi_\nu \rangle$ and $E_\nu$ are the $\nu$-th eigenstate and eigenenergy of the system, respectively ($\nu=0$ corresponds to the ground state). 

\begin{figure}
\centering
\includegraphics[width=\linewidth]{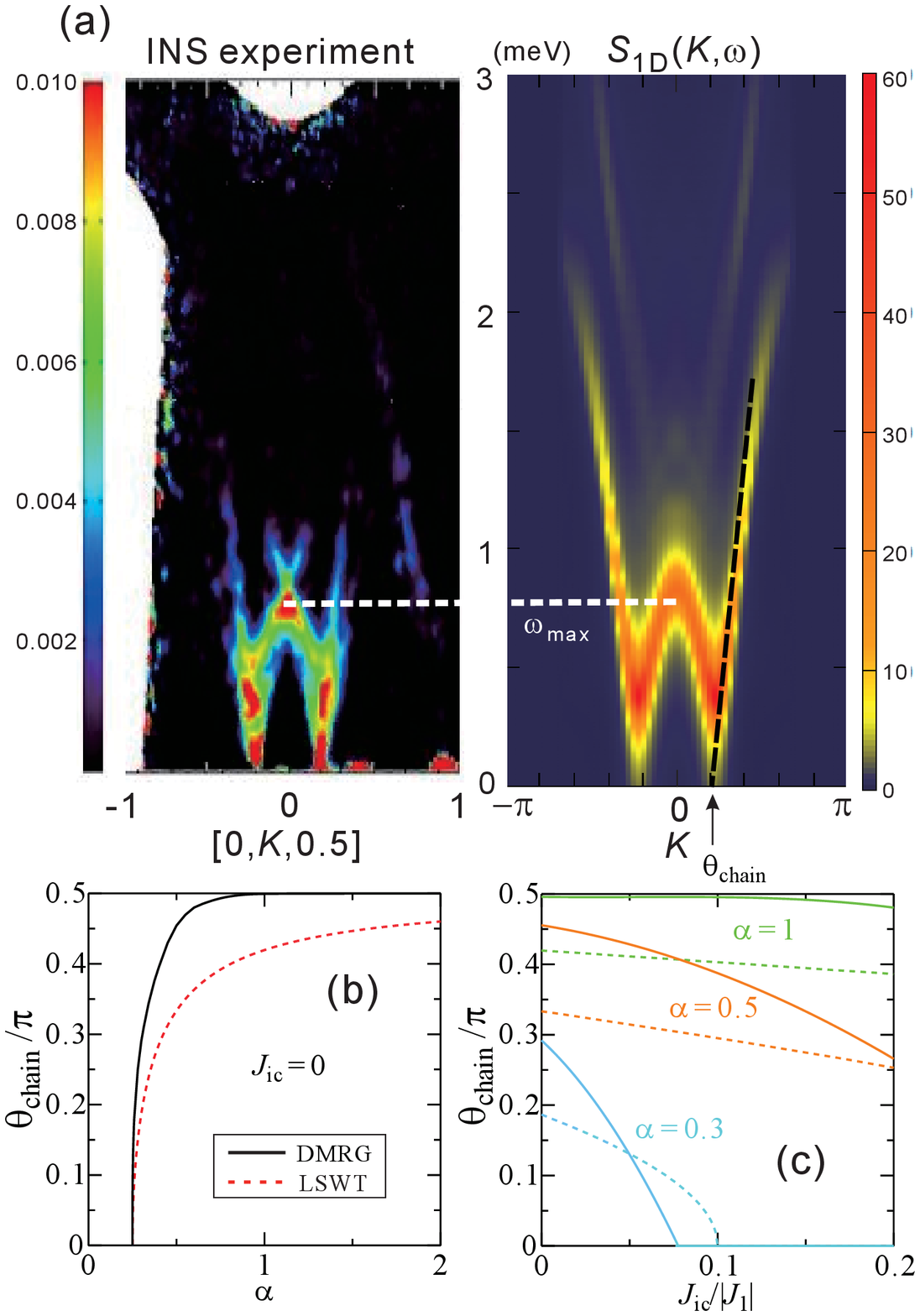}
\caption{(Color online) (a) The observed INS spectrum and the calculated dynamical spin structure factor $S_{\rm 1D}(Q,\omega)$ for $J_1 = -78$\,K, $J_2 = 28$\,K, and $J_{ic} = 7$\,K calculated by the DDMRG with $32\times2$ cluster. Comparison of pitches calculated by the DDMRG and LSWT as a function of (b) the frustration ratio $\alpha$ without interchain couplings ($J_{ic} = 0$) and (c) finite $J_{ic}/J_1$ for various values of $\alpha$.}
\label{fig_32x2}
\end{figure}

To obtain a reliable fit, we consider three quantities: i) the pitch, {\it i.e.}, the propagation vector of the spiral, $K = \theta_{\rm chain}$, ii) the maximum excitation energy at $K = 0$ in the main dispersion $\omega_{\rm cross}$, and iii) the slope of this main dispersion at $K = \theta_{\rm chain}$ ($ = d\omega/dK|_{K=\theta_{\rm chain}}$). We estimate $J_1$, $J_2$, and $J_{ic}$ to reproduce those INS experimental values; $\theta_{\rm chain} \approx 33^\circ$, $\omega_{\rm cross} \approx 0.8$\,meV, and $d\omega/dK|_{K=0.186} \approx 6.88/\pi$. As a result, we obtain $J_1 = -78$\,K, $J_2 = 28$\,K, and $J_{ic} = 7$\,K. In Fig.~\ref{fig_32x2}(a) we illustrate the good agreement of the resulting spin excitation spectrum along the chain axis with the INS data. As well, the agreement to the LSWT values is reasonably good (see Tab. \ref{tab:j}).

Given the semiquantitative agreement between DDMRG and LSWT, does it mean that LSWT can be generally a good tool to analyze INS data of frustrated systems? The answer is, that it does not work generally for a quantum spin chain system. To show this, we compare the pitches calculated by the DDMRG and LSWT. First, we set $J_{ic} = 0$. In Fig.~\ref{fig_32x2}(b) the pitch $\theta_{\rm chain}$ is plotted as a function of the frustration ratio $\alpha$. We can see that for a given $\alpha$($>1/4$) the LSWT significantly underestimates $\theta_{\rm chain}$ due to ignoring the spin fluctuations. Nevertheless, the qualitative tendency is similar and, perhaps, one might think this discrepancy is not very large; however, the quantitative discrepancy may give rise to a significant problem when we estimate the effective $\alpha$ from the observed pitch. 

In practice, let us estimate $\alpha$ from given pitches: e.g., $\alpha = 0.47$ and $1.44$ from $\theta_{\rm chain} = 80^\circ$, $\alpha = 0.33$ and $0.5$ from $\theta_{\rm chain} = 60^\circ$, and $\alpha = 0.27$ and $0.33$ from $\theta_{\rm chain} = 40^\circ$ are estimated by the DDMRG and LSWT, respectively. The agreement between both methods becomes better upon approaching $\alpha = 1/4$ (or towards small $\theta_{\rm chain}$) because then the spin fluctuations are rapidly suppressed. Whereas, a pitch analysis using the LSWT is quite difficult in case of pitches near $90^\circ$. A similar trend is also found for $J_{ic} \neq 0$. In Fig.~\ref{fig_32x2}(c) the pitches calculated by the DDMRG and LSWT are compared as a function of the interchain coupling $J_{ic}/J_1$ for fixed values of $\alpha=0.3$, $0.5$, and $1$. Again, quantitative agreement between the DDMRG and LSWT is achieved near $\alpha = 1/4$, only.

Since linarite \lin \, has a small pitch $\approx 33^\circ$ indicating a substantial suppression of the spin fluctuations, the data analysis by means of LSWT works relatively well. Thus, we conclude that the LSWT could be a good tool to (even) quantitatively analyze the INS data of frustrated chain materials when the frustration ratio is close to or less than the FM critical point $\alpha=1/4$.

In conclusion, the zero-field spin-wave excitation spectrum of linarite has been mapped out using time-of-flight inelastic neutron scattering. The low energy modes, as defined by the NNN coupling, $J_2$, and the interchain coupling, $J_{ic}$, have been clearly observed and extracted with LSWT and DDMRG. Good agreement has been found when modeling the data with the exchange interactions $J_1$, $J_2$ and $J_{ic}$, as determined from susceptibility and neutron diffraction results \citep{WillPRL2016, Wolter2012}. This way we arrived for linarite at a consistent description of the magnetic coupling, at variance to other related edge-sharing frustrated cuprate chain compounds where a consensus about the basic exchange parameters has not been achieved yet. The high energy excitations were not observed, possibly due to the weak scattering from the small sample. More detailed analysis of the experimental data by means of DDMRG highlighted the limitations inherent to the LSWT analysis and produced a set of modified $J$ values, as compared to those reported previously. Only, for linarite, with a small frustration ratio $\alpha = 0.36$ close to the FM critical point these quantitative corrections are moderate.

\begin{acknowledgments}
We are grateful for the local support staff at the SNS. This research was supported in part by the DFG under grants no. WO 1532/3-2, SU 229/9-2 and SFB-1143. This research used resources at the Spallation Neutron Source, a DOE Office of Science User Facility operated by the Oak Ridge National Laboratory. We acknowledge fruitful discussions with N. Shannon, O. Starykh and U. R\"o\ss ler. We thank G. Heide and M. G\"{a}belein from the Geoscientific Collection in Freiberg for providing the linarite crystals.
\end{acknowledgments}

\end{document}